\documentclass[sowpacs,showkeys, groupedaddress,twocolumn,
floatfix, prb, aps]{revtex4-2}
\usepackage[english]{babel}
\usepackage{fontenc}
\usepackage{graphicx}
\usepackage{amsmath}
\usepackage{epstopdf}
\usepackage{amssymb}
\usepackage{amsbsy}
\usepackage{amscd}
\usepackage{float}
\usepackage{color}
\usepackage{xcolor}
\usepackage{braket}
\usepackage{standalone}
\usepackage{bm}
\usepackage{siunitx}
\usepackage[normalem]{ulem}
\usepackage{verbatim}
\usepackage{url}
\usepackage[verbose,hypertexnames=false,bookmarksopenlevel=1,filecolor=blue,
linkcolor=blue,citecolor=blue,urlcolor=blue,pdfstartview=FitH,bookmarksopen,bookmarksnumbered,
colorlinks,plainpages=false,linktocpage]{hyperref}
\usepackage[capitalize]{cleveref}

\begin{document}
%----------------------------------------------------------------------
\title{Imperfect Many-Body Localization in Exchange-Disordered Isotropic Spin Chains}
\author{Julian Siegl and John Schliemann}
\affiliation{Institute for Theoretical Physics, University of Regensburg, Regensburg, Germany}
\date{\today}
\begin{abstract}
  We investigate many-body localization in isotropic Heisenberg spin chains with the local exchange parameters being subject to quenched disorder. Such systems incorporate a nonabelian symmetry in their Hamiltonian by an invariance under global SU(2)-rotations. Nonabelian symmetries are predicted to hinder the emergence of a many-body localized phase even in presence of strong disorder. We report on numerical studies using exact diagonalization for chains of common spin length $1/2$ and $1$. The averaged consecutive-gap ratios display a transition compatible with a crossover from an ergodic phase at small disorder strength to an incompletely localized phase at stronger disorder. The sample-to-sample variance of the averaged consecutive-gap ratio displays a maximum at the transition and distinguishes the incompletely localized phase from a fully many-body localized phase by its scaling behavior.
\end{abstract}
\maketitle
\section{Introduction \label{intro}}
\begin{figure*}[t]
    \includegraphics[scale=1]{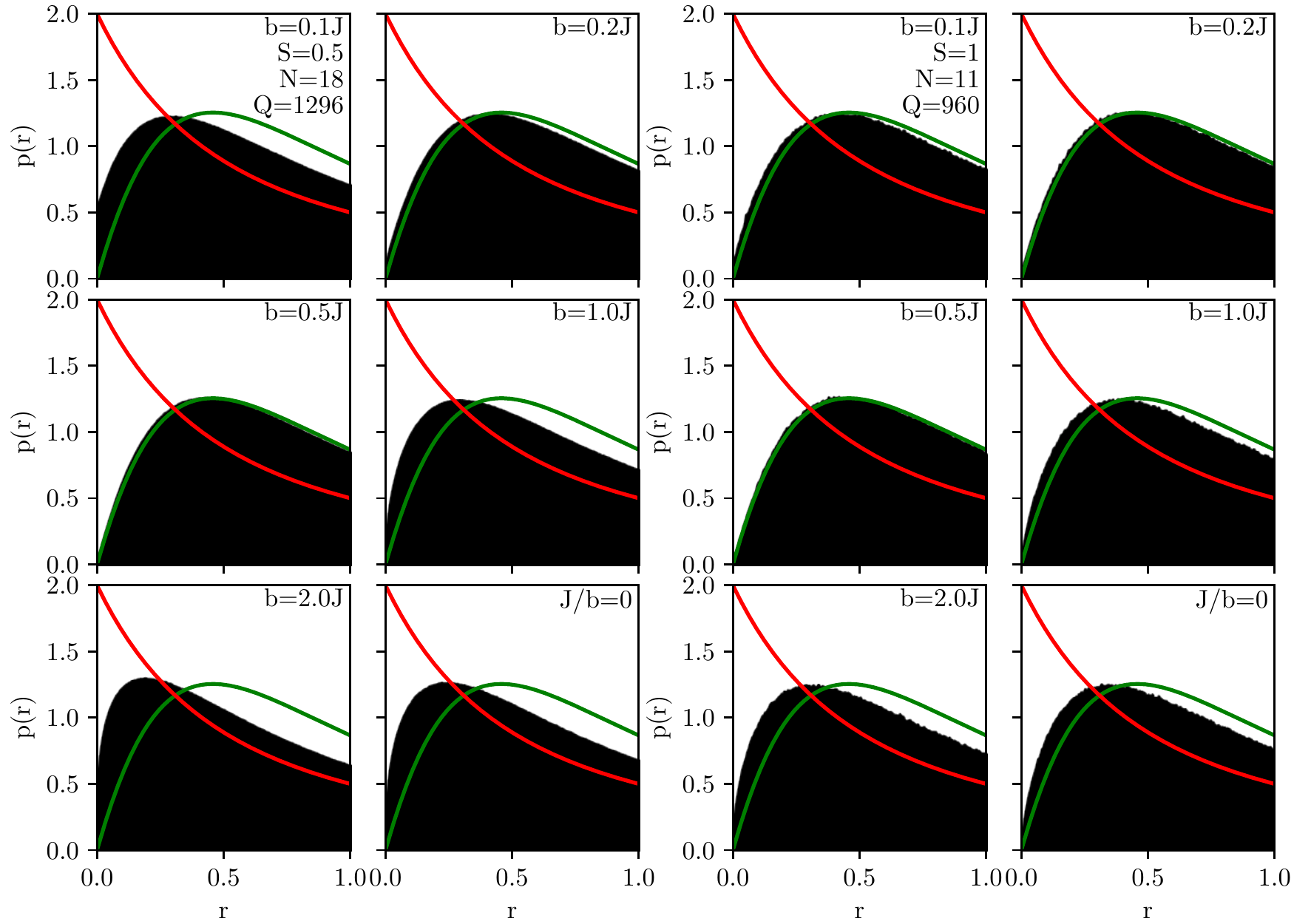}
    \caption{The probability distribution for the consecutive-gap ratio $r$ {of the multiplets} of the isotropic spin chain (\cref{ham}) with spin lengths $S=1/2$ (left) and $S=1$ (right) for different disorder strength $b$ obtained {via} exact-diagonalization. The green and the red curves show the distribution for GOE and Poisson statistics, respectively~\cite{Atas13}. The data in the bottom right panels ($J/b=0$) corresponds to infinite disorder.}
    \label{fig1}
\end{figure*}
Many-body localization amounts to the absence of thermalization in isolated interacting quantum systems in the presence of typically strong disorder~\cite{Nandkishore15, Altman15, Imbrie16a, Agarwal17, Luitz17, Haldar17, Abanin17, Abanin18}. Such a behavior is in contrast to ergodic phases where the eigenstate thermalization hypothesis is fulfilled, i.e. any appropriate subsystem of the isolated total system (being in a pure state) is accurately described by equilibrium statistical mechanics~\cite{Deutsch91, Srednicki94}.

Of particular interest is the influence of global symmetries on the competition between many-body localization and ergodicity. Potter and Vasseur~\cite{Potter16} have argued that nonabelian symmetries should generically lead to ergodic behavior due large degeneracies in the spectrum of (quasi-)local integrals of motion. Protopopov {\em et al.}~\cite{Protopopov17, Protopopov20} have performed a numerical study of disordered SU(2)-invariant spin chains and concluded the possibility of nonergodic phases different from conventional many-body localization found in other systems. Most recently, the effect of nonabelian symmetries on eigenstate thermalization~\cite{Murthy22} and entanglement entropies~\cite{Majidy22} has been discussed.

In the present work we report on a study similar to Refs.~\cite{Protopopov17, Protopopov20}, albeit different in several important aspects, of disordered rotationally invariant spin chains. A key ingredience is the sample-to-sample variance of the expectation value of the consecutive-gap ratio~\cite{Schliemann21}. As we shall see in the following, this technical tool allows to identify a novel {\em incompletely many-body localized phase}.

\section{Model and Methods \label{modelmethods}}
We study isotropic Heisenberg spin chains with disordered local exchange parameters,
\begin{equation}
  H_{\rm ex}=\sum_{i=1}^NJ_i\vec S_i\cdot\vec S_{i+1}\,,\quad J_i=J+b_i\,,
  \label{ham}
\end{equation}
with $J\geq 0$, and $b_i\in[-b,b]$ being drawn from a uniform distribution with disorder strength $b$. Periodic boundary conditions are assumed, $\vec S_{i+N}=\vec S_i$, and we consider systems with common spin length $S=1/2$ and $S=1$. As the total spin $\vec S_{\rm tot}=\sum_{i=1}^N\vec S_i$ commutes with the Hamiltonian, all energy levels lie in multiplets of $\vec S_{\rm tot}$. For our numerical analysis based on exact diagonalization we use all multiplets except for those with quantum numbers $S_{\rm tot}\in\{NS,NS-1\}$.

{For disorder strengths $b\geq J$ the exchange parameters $J_i$ can for some links approach zero. If at least two such ``weak links'' occur the chain splits effectively into two separate systems. Localization in this sense is distinct from the dynamic localization characterizing the many-body localized phase. Therefore, we briefly estimate the importance of the aforementioned mechanism for the subsequent results. For the uniform distribution considered here, the expectation value of the weakest link~\cite{Protopopov20} $q:=\min\{|J_1|,\dots,|J_N|\}$ is, for $b\leq J$, given by
\begin{equation}
  \langle q\rangle=J-\frac{N-1}{N+1}b\,,
  \label{hamwlink1}
\end{equation}
whereas for $b\geq J$ one obtains
\begin{equation}
  \langle q\rangle=\frac{b}{N+1}+\frac{J}{N+1}\left(\frac{J}{b}\right)^N\,,
  \label{hamwlink2}
\end{equation}
as derived in \cref{WeakestLink}. The influence this weakest link has on the connectivity of the chain can be estimated from the ratio between $q$ and the dominant energy scale of the Hamiltonian $\max\{J,b\}$. Defining $\tilde {q}=q/\max\{J,b\}$, we find that $\langle \tilde{q} \rangle$ approaches its lowest value $1/(N+1)$ as $(J/b)\rightarrow 0$. Even there, $\langle \tilde{q}\rangle$ decays only with the inverse of the system size $N$. For low $N\leq18$ as considered in this work, this minimum value still is larger than the mean level spacing even for the lowest dimensionality $S_{\rm tot}$ subspace of the Hamiltonian. We conclude, that for the chain lengths considered throughout this work the above mentioned possible fragmentation of the system remains unlikely.}

Spectral properties of the Hamiltonian (\cref{ham}) are investigated via the consecutive-gap ratio $r$~\cite{Oganesyan07,Atas13} defined as
\begin{equation}
  r_n=\frac{\min\{s_n,s_{n-1}\}}{\max\{s_n,s_{n-1}\}}\,,
  \label{ratgap1}
\end{equation}
where $s_n=e_{n+1}-e_n$ is the difference between neighboring energy levels $e_{n+1}$, $e_n$, each corresponding to a multiplet of given total spin quantum number $S_{\rm tot}$. {The average $\langle r\rangle=\int_0^1drp(r)r$ of the random variable $r=r_n$ is determined by its probability distribution $p(r)$. Importantly, different phases reflect via the associated distributions in characteristic values for this averaged consecutive-gap ratio, making it a useful tool to elucidate phase transitions from the spectral information about a system. For integrable phases like the many-body localized phase, the level spacing follows Poissonian statistics, which gives $\langle r \rangle=2\ln 2-1\approx 0.3863$~\cite{Abanin17}. Contrary, random matrix theory predicts $\langle r \rangle=4-2\sqrt{3}\approx 0.5359$ for the Gaussian orthogonal ensemble (GOE) connected to the ergodic phase with the symmetries of the Hamiltonian considered here~\cite{Atas13}.}

To calculate the average, we perform calculations for a large number $Q\gg 1$ of disorder realizations. $\langle r \rangle$ can then be approximated by first calculating the average $\langle r \rangle_{\alpha}$ for each disorder realization (or sample) $\alpha\in\{1,\dots,Q\}$ and then averaging the results,
\begin{equation}
  \langle r\rangle
  =\lim_{Q\to\infty}\frac{1}{Q}\sum_{\alpha=1}^{Q}\langle r \rangle_{\alpha}
  \approx\frac{1}{Q}\sum_{\alpha=1}^{Q}\langle r \rangle_{\alpha}
  \quad,\quad
  Q\gg 1\,.
  \label{ratgap2}
\end{equation}
The random variable $x=\langle r\rangle_{\alpha}$ follows the distribution
\begin{eqnarray}
  s(x) & = & \frac{1}{(2b)^N}\int_{-b}^b db_1\cdots\int_{-b}^b db_N\nonumber\\
  & & \qquad\cdot\delta\left(x-\int_0^1drp(r;b_1,\dots,b_N)r\right)\,,
  \label{ratgap2a}
\end{eqnarray}
where $p(r;b_1,\dots,b_N)=p_{\alpha}$ is the probability distribution within a system with local modulations $b_1,\dots,b_N$ on the exchange strength forming the disorder realization $\alpha$.

{Performing this step wise averaging yields additional information, since next to its average $\langle r \rangle$, $s(x)$ also contains information on the sample-to-sample variance, i.e. the variance of $\langle r\rangle_{\alpha}$ with respect to the probability distribution (\cref{ratgap2a}). The latter has been used in previous works~\cite{Schliemann21} as an additional tool indicating the presence of a phase transition.}

The sample-to-sample variance can be approximated as
\begin{equation}
  (\Delta_s r)^2\approx\frac{1}{Q}\sum_{\alpha=1}^Q
  {(\langle r\rangle_{\alpha}-\langle r\rangle)}^2\,,
  \label{ratgap3}
\end{equation}
and is to be distinguished from the total variance $(\Delta_p r)^2$ with respect to the probability distribution $p(r)$. Indeed, these two quantities are related via~\cite{Schliemann21}
\begin{equation}
  (\Delta_p r)^2=(\Delta_s r)^2
  +\lim_{Q\to\infty}\frac{1}{Q}\sum_{\alpha=1}^{Q}(\Delta_{\alpha} r)^2\,,
  \label{ratgap4}
\end{equation}
where $(\Delta_{\alpha} r)^2$ is the variance within the disorder realization $\alpha$. For all further technical details we refer to Ref.~\cite{Schliemann21}.

We will contrast our results for the Hamiltonian (\cref{ham}) with previous findings for a Heisenberg chain being subject to a disordered local field coupling to the $z$-component of each spin,
\begin{equation}
  H_{\rm lf}=J\sum_{i=1}^N\vec S_i\cdot\vec S_{i+1}
  +2S\sum_{i=1}^Nh_iS^z_i\,,
  \label{hamlf}
\end{equation}
with uniformly distributed local field strengths $h_i\in[-h,h]$. This model is, mainly for spin length $S=1/2$, commonly used in studies of many-body localization~\cite{Nandkishore15, Altman15, Imbrie16a, Agarwal17, Luitz17, Haldar17, Abanin17, Abanin18}. Fewer works considered also larger spin lengths~\cite{Gu19, Richter20, Schliemann21}. The system shows a transition from an ergodic phase at small disorder strength $h$ to a many-body localized phase at large disorder. Numerical studies~\cite{Pal10, Luitz15, Devakul15, Doggen18, Chanda20a} have placed the critical disorder strength at values of $h_{\rm cr}\approx 4J$ or larger. The nature and position of this transition has been critically discussed in recent publications in terms of matrix product states~\cite{Doggen21}, Renyi entropies~\cite{Kiefer21}, quantum avalanches~\cite{Morningstar22}, Liouvillian relaxation~\cite{Sels22}, and time evolution~\cite{Sierant22}. On the other hand, investigations in Ref.~\cite{Schliemann21} found, based on the sample-to-sample variance, a transition at $h_{\rm cr}\approx 2.6J\dots 3.0J$ for $S=1/2$, and $h_{\rm cr}\approx 4.0J\dots 4.5J$ for $S=1$.

\section{Results \label{results}}

\begin{figure*}[t]
    \includegraphics[scale=1]{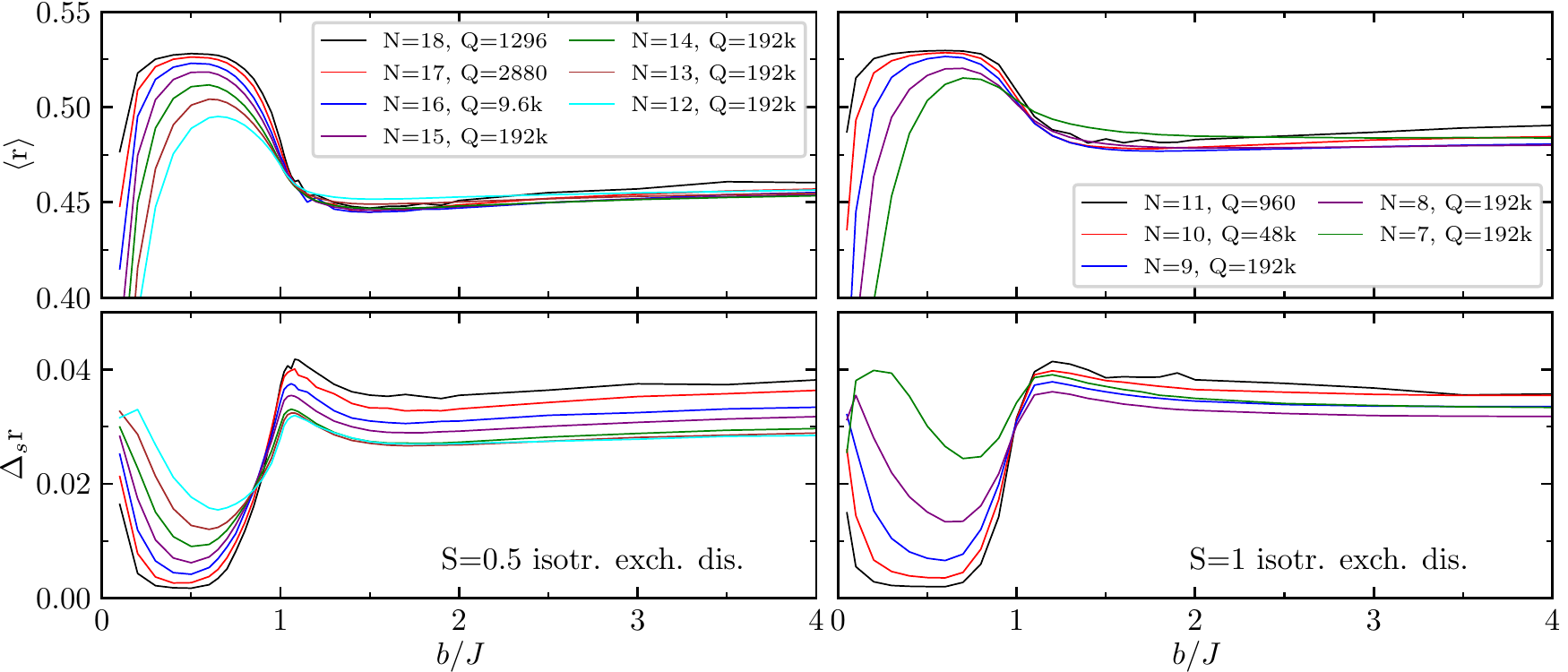}
    \caption{Top panels: The expectation value $\langle r\rangle$ as a function of disorder strength in chains of spin length $S=1/2$ (left) and $S=1$ (right) for spin chains of finite sizes $N$ and pertaining numbers $Q$ of disorder realizations. Bottom panels: The sample-to-sample standard deviation $\Delta_s r$ for the same parameters as in the top panels.}
    \label{fig2}
\end{figure*}

\cref{fig1} shows the disorder-averaged probability distribution $p(r)$ of the isotropic spin chain (\cref{ham}) in finite systems for spin lengths $S\in\{1/2,1\}$. For small disorder the system is close to an ergodic phase {with the residual difference being due to the approximate translational invariance for weak disorder. This symmetry results in exact degeneracy of the multiplets for vanishing disorder but gets lifted as soon as the local nature of the disorder becomes appreciable. At intermediate disorder} $b=:b_{\rm erg}\approx 0.5J$ the data is practically indistinguishable from the GOE prediction. {Importantly, with increasing disorder $p(r)$ does not remain aligned with the GOE prediction but shifts towards smaller values in a fashion similar to the transient regime observed before the onset of many-body localization in systems of the form of \cref{hamlf}. Contrary to this transition, which has been thoroughly been investigated in literature, the distribution for the isotopically exchange-disordered spin chain does not reach, even for infinite disorder, Poissonian statistics. This is in line with the argument of Potter and Vasseur~\cite{Potter16} which rules out the possibility of the emergence of a fully integrable phase due to the nonabelian symmetry incorporated in the Hamiltonian in \cref{ham}.}

\begin{figure}[b]
  \includegraphics[scale=1]{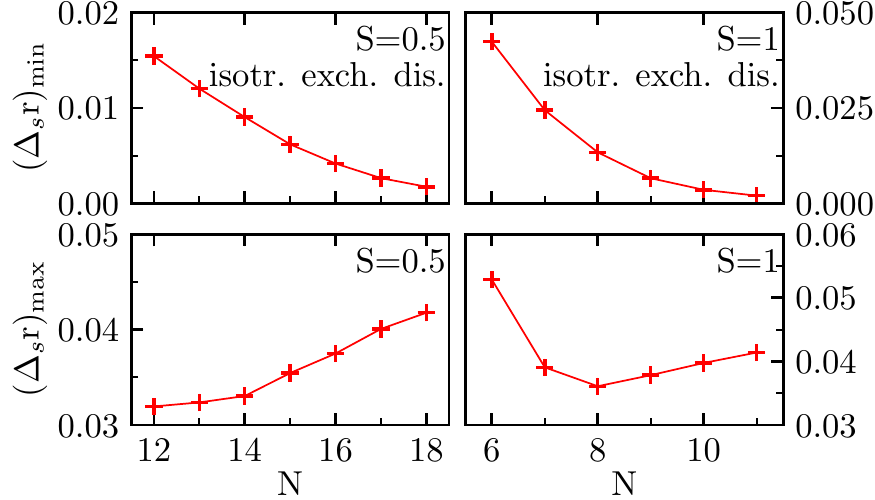}
  \caption{The minimum ${(\Delta_s r)_{\rm min}}$ (top panels) and the maximum ${(\Delta_s r)_{\rm max}}$ (bottom panels) of the sample-to-sample standard deviation for spin length $S=1/2$ (left) and $S=1$ (right) as a function of system size $N$.}
  \label{fig3}
\end{figure}

\begin{figure}[b]
 \includegraphics[scale=1]{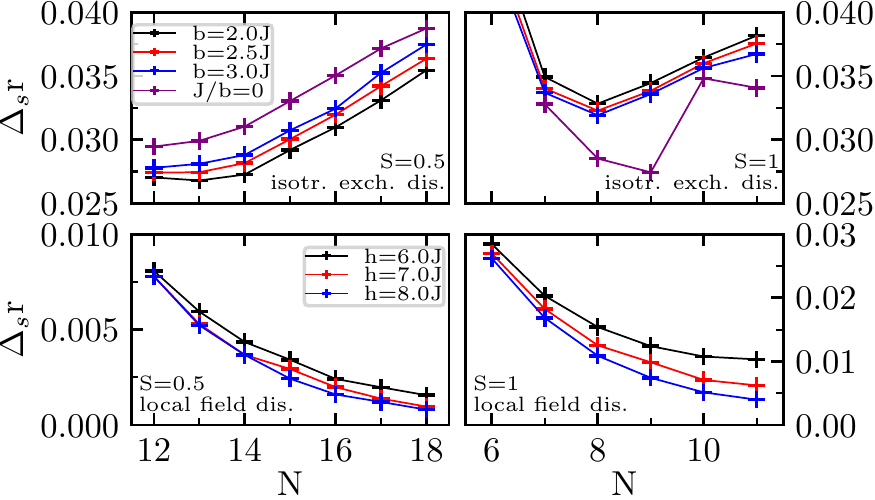}
  \caption{The sample-to-sample standard deviation $\Delta_s r$ outside the ergodic phase for the isotropic disordered Heisenberg chain (\cref{ham}) (top panels) and the chain (\cref{hamlf}) subject to a local disordered field (bottom panels). The data in the bottom panels is adopted from Ref.~\cite{Schliemann21}.}
  \label{fig4}
\end{figure}
\begin{figure*}[t]
  \includegraphics[scale=1]{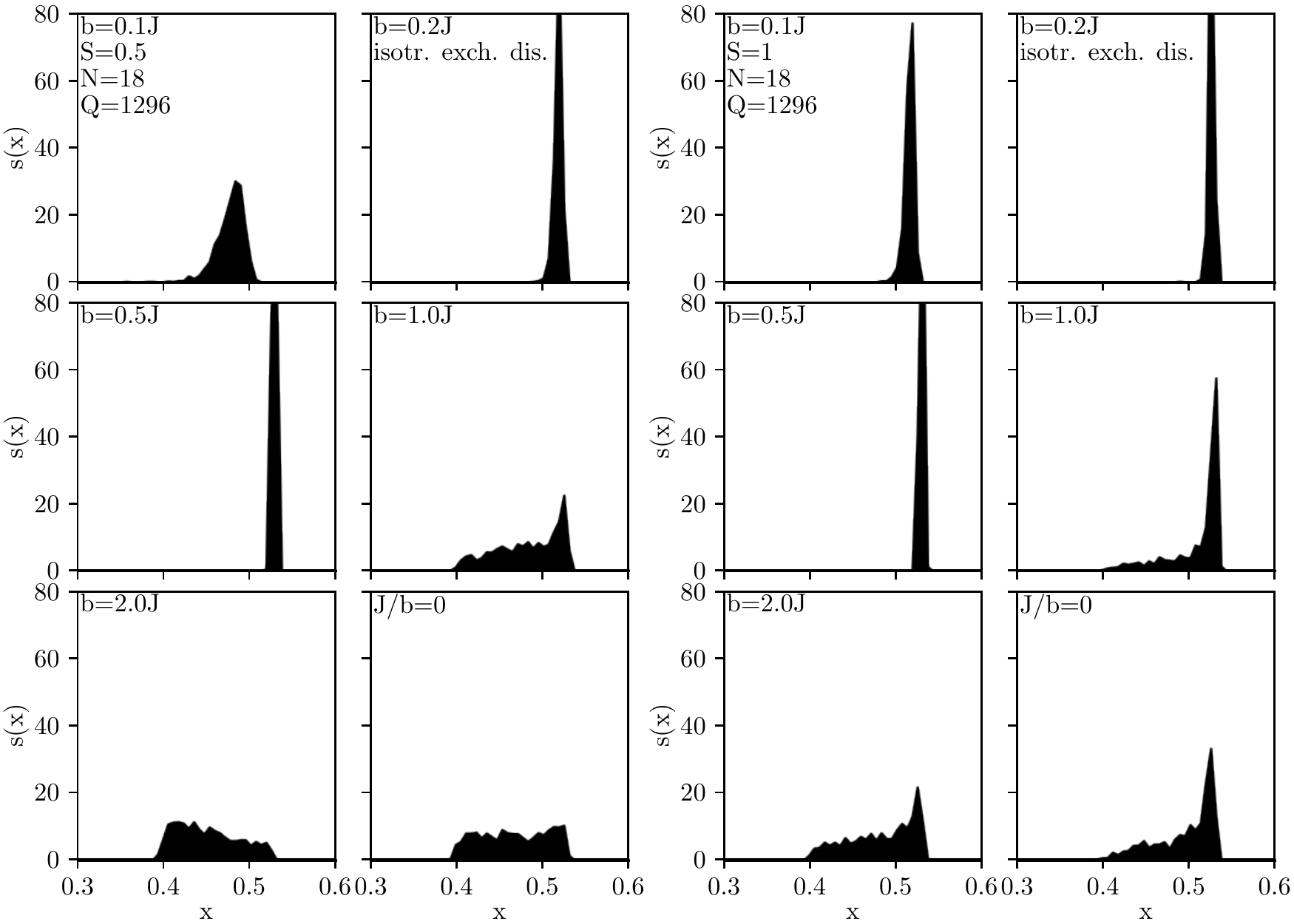}
  \caption{The probability distribution (\cref{ratgap2a}) for the realization-specific average $\langle r\rangle_{\alpha}$ for disordered Heisenberg chains of spin length $S=1/2$ (left) and $S=1$ (right). The disorder strengths $b$ in both panels are the same as in \cref{fig1}.}
  \label{fig5}
\end{figure*}

In \cref{fig2} we display the expectation value $\langle r\rangle$ along with the sample-to-sample standard deviation $\Delta_s r$ as a function of disorder strength $b$ for spin chains of different length. Consistently with the findings from \cref{fig1}, the expectation value of the consecutive-gap ratio has a maximum at $b=b_{\rm erg}\approx 0.5J$, which moves with increasing system size close to the GOE value $\langle r \rangle\approx 0.5359$. At the same disorder strength the sample-to-sample standard deviation reaches a minimum $(\Delta_s r)_{\rm min}$, and at larger $b=:b_{\rm cr}\gtrsim 1.0J$, a maximum $(\Delta_s r)_{\rm max}$ is attained while $\langle r\rangle$ shows an inflection point at about the same disorder strength.
{Note that the transition starts already at $b/J$ below the critical value of 1, at which weak links may occur. Together with the previous discussion on their low probability of occurring, we thus rule out a fragmentation of the chain due to weak links as the root cause of the observed transition. Especially the observed inflection point is not susceptible to an interpretation in terms of a fragmentation since $\langle \tilde q \rangle$ monotonously decreases above $b/J=1$.}

The behavior of $(\Delta_s r)_{\rm min}$ and ${(\Delta_s r)_{\rm max}}$ as a function of system size $N$ in shown in \cref{fig3}. As seen there, ${(\Delta_s r)_{\rm min}}$ decreases monotonously, and the finite-size data is consistent with the assumption $(\Delta_s r)_{\rm min}\to 0$ for $N\to\infty$, while $(\Delta_s r)_{\rm max}$ increases with $N$. Thus, viewing the expectation value $\langle r\rangle$ as an order parameter, this quantity, along with the standard deviation $\Delta_s r$, shows as a function of disorder strength typical features of a phase transition. Analogous observation were made for the anisotropic chain (\cref{hamlf}) being subject to a local disordered field~\cite{Schliemann21}.

Let us now analyze the phase of the isotropic exchange-disordered spin chain (\cref{ham}) for disorder strengths larger than the critical value $b_{\rm cr}\gtrsim 1.0J$. As the top panels of \cref{fig2} show, the average $\langle r\rangle$ of the consecutive-gap ratio lies here, for the system sizes $N$ considered, between the value for Poissonian statistics and GOE statistics. However, $\langle r\rangle$ increases slightly with $N$, and a meaningful extrapolation turns out to be problematic; similar observations were made in Ref.\cite{Protopopov20,note1}. Thus, just from the finite-size data for $\langle r\rangle$ at hand, it is difficult to argue that the system behaves in the thermodynamic limit $N\to\infty$ differently from a GOE ergodic phase.

A qualitative difference between the level statistics for $b>b_{\rm cr}\gtrsim 1.0J$ is, however, revealed by the sample-to-sample standard deviation $\Delta_s r$. Let us first discuss the case of spin length $S=1/2$. As displayed in the top left panel of \cref{fig4}, $\Delta_s r$ increases, at given disorder strength $b>b_{\rm cr}$, with system size (which is also seen in the bottom left panel of \cref{fig2}). This behavior is in contrast to the ergodic phase around $b=b_{\rm erg}\approx 0.5J$ where $\Delta_s r$ decreases with growing $N$ (see also the top panels of \cref{fig3}), and the same is observed in the ergodic phase of the model (\cref{hamlf}) lacking a nonabelian symmetry~\cite{Schliemann21}. Finally, in the bottom left panel of \cref{fig4} we have additionally plotted $\Delta_s r$ for the anisotropic Hamiltonian (\cref{hamlf}) at large disorder (where the system is many-body localized), and clearly the data also decreases with system size.

As of the case $S=1$, $\Delta_s r$ as a function of $N$ plotted in the top right panel of \cref{fig4} behaves different from the ergodic phase, but with less clear finite-size behavior. Moreover this quantity decreases with growing disorder strength $b$, which is in contrast to $S=1/2$ where the data smoothly increases with disorder strength.

In summary, the isotropic exchange-disordered spin chain (\cref{ham}) undergoes at a critical disorder strength $b_{\rm cr}\gtrsim 1.0J$ a transition from an ergodic phase to a phase which is neither ergodic nor fully many-body localized. {As the nonergodic nature of this phase, even at strong disorder, is similar to the transient regime of the local field disordered chain (\cref{hamlf}) before integrability is reached, we refer to it as the {\em incompletely many-body localized phase}.} %We therefore refer to this phenomenon as {\em incomplete many-body localization}.

The novel character of the incompletely many-body localized phase is also reflected by the probability distribution (\cref{ratgap2a}) for the realization-specific average $\langle r\rangle_{\alpha}$. We plot it in \cref{fig5} for the same representative values of the disorder strength as in \cref{fig1}. Starting from weak disorder, the distribution $s(x)$ narrows to a peak close to the GOE value $\langle r \rangle\approx 0.5359$ when approaching $b=b_{\rm erg}\approx 0.5J$. At the transition, $b=:b_{\rm cr}\gtrsim 1.0J$, $s(x)$ broadens, and remains of similar width for larger disorder. The latter observation is in qualitative contrast to the model (\cref{hamlf}) at strong disorder, where the distribution again narrows to a peak~\cite{Schliemann21}.

\section{Summary and outlook}
\label{sumout}
{Using methods developed in previous works~\cite{Schliemann21}, we studied the spectra of isotropic exchange-disordered spin chains (\cref{ham}) taking into account the probability distribution across samples. The latter reveals a transition from the ergodic phase at low to intermediate disorder strength to a phase qualitatively distinct from both the ergodic as well as the many-body localized phase. Such a phase has been theorized to result as a consequence of nonabelian symmetries, as present in \cref{ham}, disrupting the formation of a sufficiently exhaustive set of local integrals of motion as required by many-body localization~\cite{Potter16,Abanin17}.} This novel phase is distinguished from a fully many-body localized phase as well as from the ergodic phase by the scaling behavior of the sample-to-sample standard deviation of the consecutive-gap ratio with system size. Moreover, our numerical data shows a qualitative difference between the spin lengths $S=1/2$ and $S=1$. This is in contrast to the local field disordered spin model (\cref{hamlf}) { where no such qualitative difference between spin lengths has been observed}~\cite{Schliemann21}. Further differences between the incompletely many-body localized phase and the fully many-body localized phase are found in the probability distribution for the realization-specific average of the consecutive-gap ratio. {Beyond the spectral information investigated here, future research may strife to use complimentary numerical methods to verify the presence of this phase and elucidate its properties.}

\begin{acknowledgments}
  We thank J. Richter for useful discussions and P. Wenk for technical help with the numerics. The work of J. Siegl was supported by Deutsche Forschungsgemeinschaft via SFB 1277.
\end{acknowledgments}

\appendix
\section{Derivation of the Expectation Value of the Weakest Link}
\label{WeakestLink}

The probability distribution for the random variable $v=|J_i|$ with $J_i$ given in \cref{ham} is $\pi(v)=1/(2b)$ for $b\leq J$, and for $b\geq J$
\begin{equation}
  \pi(v)=\left\{
    \begin{array}{cc}
      1/b & v\in[0,b-J] \\
      1/(2b) & v\in[b-J,b+J]
    \end{array}\right.\,.
    \label{wlink1}
\end{equation}
Moreover, the probability for one coupling in the Hamiltonian (\cref{ham}) having modulus $|J_i|=q$ while all other $N-1$ moduli are larger is
\begin{equation}
  \Pi(q)=N\pi(q)\left(\int_q^{b+J}dv\pi(v)\right)^{N-1}\,,
  \label{wlink2}
\end{equation}
which reads explicitly for $b\leq J$
\begin{equation}
  \Pi(q)=\frac{N}{2b}\left(\frac{b+J-q}{2b}\right)^{N-1}\,,
  \label{wlink3}
\end{equation}
and for $b\geq J$
\begin{equation}
  \Pi(q)=\left\{
    \begin{array}{cc}
      \frac{N}{b}\left(\frac{b-q}{b}\right)^{N-1} & q\in[0,b-J] \\
      \frac{N}{2b}\left(\frac{b+J-q}{2b}\right)^{N-1} & q\in[b-J,b+J]
    \end{array}\right.\,.
    \label{wlink4}
\end{equation}
From these distributions the expectation value of $q$ follows as given in \cref{hamwlink1,hamwlink2}.
%----------------------------------------------------------------------


\begin{thebibliography}{10}
%----------------------------------------------------------------------

\bibitem{Nandkishore15}
  R. Nandkishore and D.~A. Huse,
  %{\em Many-Body Localization and Thermalization
  %  in Quantum Statistical Mechanics},
  Ann. Rev. Cond. Mat. Phys. {\bf 6}, 15 (2015).

\bibitem{Altman15}
  E. Altman and R. Vosk,
  %{\em Universal Dynamics and Renormalization in Many-Body-Localized
  %  Systems},
  Ann. Rev. Cond. Mat. Phys. {\bf 6}, 383 (2015).

\bibitem{Imbrie16a}
  J.~Z. Imbrie, V. Ros, and A. Scardicchio,
  %{\em Local Integrals of Motion in Many-Body Localized systems},
  Ann. Phys. (Berlin) {\bf 529}, 1600278 (2017).

\bibitem{Agarwal17}
  K. Agarwal, E. Altma, E. Demler, S. Gopalakrishnan, D.~A. Huse, and M. Knap,
  %{\em Rare-region effects and dynamics near
  %  the many-body localization transition},
  Ann. Phys. (Berlin) {\bf 529}, 1600326 (2017).

\bibitem{Luitz17}
  D.~J. Luitz and Y. Bar Lev,
  %{\em The ergodic side of the many-body localization transition},
  Ann. Phys. (Berlin) {\bf 529}, 1600350 (2017).

\bibitem{Haldar17}
  A. Haldar and A. Das,
  %{\em Dynamical Many-body Localization and Delocalization
  %  in Periodically Driven Closed Quantum Systems},
  Ann. Phys. (Berlin) {\bf 529}, 1600333 (2017).

\bibitem{Abanin17}
  D.~A. Abanin and Z. Papic,
  %{\em Recent progress in many-body localization},
  Ann. Phys. (Berlin) {\bf 529} , 1700169 (2017).

\bibitem{Abanin18}
  D.~A. Abanin, E. Altman, I. Bloch, and M. Serbyn,
  %{\em Many-body localization, thermalization, and entanglement},
  Rev. Mod. Phys. {\bf 91}, 021001 (2019).

\bibitem{Deutsch91}
  J.~M. Deutsch,
  %{\em Quantum statistical mechanics in a closed system},
  Phys. Rev. A {\bf 43}, 2046 (1991).

\bibitem{Srednicki94}
  M. Srednicki,
  %{\em Chaos and quantum thermalization},
  Phys. Rev. E {\bf 50} 888 (1994).

\bibitem{Potter16}
  A.~C. Potter and R. Vasseur,
  %{\em Symmetry constraints on many-body localization},
  Phys. Rev. B {\bf 94}, 224206 (2016).

\bibitem{Protopopov17}
  I.~V. Protopopov, W.~W. Ho, and D.~A. Abanin,
  %{\em Effect of SU(2) symmetry on many-body localization and thermalization},
  Phys. Rev. B {\bf 96}, 041122 (2017).

\bibitem{Protopopov20}
  I.~V. Protopopov, R.~K. Panda, T. Parolini, A. Scardicchio, E. Demler,
  and D.~A. Abanin,
  %{\em Non-Abelian Symmetries and Disorder: A Broad Nonergodic Regime
  %  and Anomalous Thermalization},
  Phys. Rev. X {\bf 10}, 011025 (2020).

\bibitem{Murthy22}
  C. Murthy, A. Babakhani, F. Iniguez, M. Srednicki, and N. Yunger Halpern,
  %{\em Non-Abelian eigenstate thermalization hypothesis},
  arXiv:2206.05310 [quant-ph].

\bibitem{Majidy22}
  S. Majidy, A. Lasek, D.~A. Huse, and N. Yunger Halpern,
  %{\em Non-Abelian symmetry can increase entanglement entropy},
  Phys. Rev. B {\bf 107}, 045102 (2023).

\bibitem{Schliemann21}
  J Schliemann, J.~V.~I. Costa, P. Wenk, and J.~C. Egues,
  %{\em Many-body localization: Transitions in spin models},
  Phys. Rev. B {\bf 103}, 174203 (2021).

\bibitem{Oganesyan07}
  V. Oganesyan and D.~A. Huse,
  %{\em Localization of interacting fermions at high temperature},
  Phys. Rev. B {\bf 75}, 155111 (2007).

\bibitem{Atas13}
  Y.~Y. Atas, E. Bogomolny, O. Giraud, and G. Roux,
  %{\em Distribution of the Ratio of Consecutive Level Spacings in Random
  %  Matrix Ensembles},
  Phys. Rev. Lett. {\bf 110}, 084101 (2013).

\bibitem{Gu19}
  J. Gu, S. Liu, M. Yazback, H.-P. Cheng, and X.-G. Zhang,
  %{\em Many-body localization from random magnetic anisotropy},
  Phys. Rev. Res. {\bf 1}, 033183 (2019).

\bibitem{Richter20}
  J. Richter, D. Schubert, and R. Steinigeweg,
  %{\em Decay of spin-spin correlations in disordered quantum and classical
  %spin chains},
  Phys. Rev. Res. {\bf 2}, 013130 (2020).

\bibitem{Pal10}
  A. Pal and D.~A. Huse,
  %{\em Many-body localization phase transition},
  Phys. Rev. B. {\bf 82}, 174411 (2010).

\bibitem{Luitz15}
  D.~J. Luitz, N. Laflorencie, and F. Alet,
  %{\em Many-body localization edge in the random-field Heisenberg chain},
  Phys. Rev. B {\bf 91}, 081103 (2015).

\bibitem{Devakul15}
  T. Devakul and R.~R.~P. Singh,
  %{\em Early breakdown of area-law entanglement at
  %  many-body delocalization transition},
  Phys. Rev. Lett. {\bf 115}, 187201 (2015).

\bibitem{Doggen18}
  E.~V.~H. Doggen, F. Schindler, K.~S. Tikhonov, A.~D. Mirlin, T. Neupert,
  D.~G. Polyakov, and I.~V. Gornyi,
  %{\em Many-body localization and delocalization in large quantum chains},
  Phys. Rev. B {\bf 98}, 174202 (2018).

\bibitem{Chanda20a}
  T. Chanda, P. Sierant, and J. Zakrzewski,
  % {\em Time dynamics with matrix product states:
  %  Many-body localization transition of large systems revisited},
  Phys. Rev. B {\bf 101}, 035148 (2020).

\bibitem{Doggen21}
  E.~V.~H. Doggen, I.~V. Gornyi, A.~D. Mirlin, and D.~G. Polyakov,
  %{\em Many-body localization in large systems: Matrix-product-state approach},
  Ann. Phys. {\bf 435}, 168437 (2021).

\bibitem{Kiefer21}
  M. Kiefer-Emmanouilidis, R. Unanyan, M. Fleischhauer, and J. Sirker,
  %{\em Slow delocalization of particles in many-body localized phases},
  Phys. Rev. B {\bf 103}, 024203 (2021).

\bibitem{Morningstar22}
  A. Morningstar, L. Colmenarez, V. Khemani, D.~J. Luitz, and D.~A. Huse,
  %{\em Avalanches and many-body resonances in many-body localized systems},
  Phys. Rev. B {\bf 105}, 174205 (2022).

\bibitem{Sierant22}
  P. Sierant and J. Zakrzewski,
  %{\em Challenges to observation of many-body localization},
  Phys. Rev. B {\bf 105}, 224203 (2022).

\bibitem{Sels22}
  D. Sels,
  %{\em Bath-induced delocalization in interacting disordered spin chains},
  Phys. Rev. B {\bf106}, L020202 (2022).

\bibitem{note1}
  See Fig.~6 of Ref.~\cite{Protopopov20}. The parameter value $\alpha=1.0$
  there corresponds to a uniform disorder probabilty distribution as in our
  Hamiltonian (\ref{ham}) with $J=0$.

\end{thebibliography}
\end{document}